\documentclass[a4paper,UKenglish,cleveref,autoref,thm-restate,
runningheads,orivec]{llncs}

\usepackage[dvipsnames]{xcolor}
\usepackage{todonotes}
\usepackage{paralist}
\usepackage{amssymb}
\usepackage{amsmath}
\usepackage[a4paper, margin=2.5cm]{geometry}
\usepackage{apxproof} 
\usepackage{pgf}
\usepackage{tikz}
\usetikzlibrary{arrows,shapes,automata,petri}

\newif\iflisten\listentrue

\newtheoremrep{theorem}{Theorem}
\newtheoremrep{lemma}{Lemma}
\usepackage{amsmath,xspace}
\usepackage{etex}
\usepackage{centernot}
\usepackage{stmaryrd}
\usepackage{wrapfig}
\usepackage{amsfonts}
\usepackage{mathtools}
\usepackage{xspace}
\usepackage{color}
\definecolor{lgray}{gray}{0.9}
\usepackage{fixltx2e}
\usepackage{fancybox}
\usepackage{ifthen}
\usepackage{pgf}
\usepackage{tikz}
\usetikzlibrary{arrows,automata}
\usepackage{graphics} 
\usepackage{subfig}
\usepackage{proof}
\usepackage{hyperref}
\usepackage{soul}
\usepackage{accsupp} 
\usepackage{cancel}
\usepackage{textcomp}
\usepackage{calrsfs}
\usepackage{paralist}
\usepackage{enumitem}

\usepackage{changebar}
\usepackage{todonotes}
\usepackage[super]{nth}

\DeclareMathAlphabet{\pazocal}{OMS}{zplmf}{m}{n}
\newcommand{\mcal}[1]{\pazocal{#1}}

\newcommand{\rulename}[1]{\mbox{\textsc{#1}}}

\newcommand{\qt}[1]{``{#1}"}

\newcommand{\rTo}[1]{\xrightarrow{#1}}

\newcommand{\true}{{\sf true}}

\newlength{\arrow}
\newcounter{sqindex}

 \newcommand{\rom}[1]{ \textup{(\lowercase\expandafter{\romannumeral#1})}}
 
\usepackage[ruled]{algorithm2e} 

\SetAlFnt{\small}
\SetAlCapFnt{\small}
\SetAlCapNameFnt{\small}
\SetAlCapHSkip{0pt}
\IncMargin{-\parindent}

\newcommand \Until      {\mathbin{\mcal{U}\kern-.1em}}
\newcommand \Release     {\mathbin{\mcal{R}\kern-.1em}}

\newcommand \Since      {\mathbin{\mcal{S}\kern-.08em}}

\newcommand \g    {\mathsf{{G}\kern.08em}}
\newcommand \f    {\mathsf{{F}\kern.08em}}
\newcommand \UntilHat   {\mathbin{\LTLhat{\mcal{U}}\kern-.1em}}

\newcommand \SinceHat   {\mathbin{\LTLhat{\mcal{S}}\kern-.08em}}

\renewcommand \models   {\mathrel{\raisebox{-.1em}{$\vDash$}}}

\newcommand \smodels    {\models\kern-.5em\models}
\renewcommand \phi      {\varphi}

\newcommand \ltl        {\textsc{ltl}\xspace}

\newcommand \ltal       {\textsc{ltol}\xspace}

\newcommand \pspace       {\textsc{pspace}\xspace}
\newcommand \expspace       {\textsc{expspace}\xspace}

\newcommand{\set}[1]{\{{#1}\}}
\newcommand{\conf}[1]{\langle{#1}\rangle}

\newcommand{\band}[3]{\bigwedge\limits_{#1}^{#2}{#3}}
\newcommand{\bor}[3]{\bigvee\limits_{#1}^{#2}{#3}}
\newcommand{\bcup}[3]{\bigcup\limits_{#1}^{#2}{#3}}
\newcommand{\bpcup}[2]{\biguplus\limits_{#1}{#2}}
\newcommand{\func}[3]{{#1}:{#2}\rightarrow{#3}}
\newcommand{\pfunc}[3]{{#1}^{#2}_{#3}}
\newcommand{\such}{{\bf\mathsf{s.t.}}\ }
\newcommand{\toall}{\star}
\newcommand{\cstate}[2]{{s}^{#1}_{#2}}
\newcommand{\sstate}[1]{{s}_{#1}}
\newcommand{\vardom}[1]{\mathsf{Dom}({#1})}
\newcommand{\vdom}{\vardom{v}}
\newcommand{\compdom}[1]{\prod_{v\in V_{#1}}\vdom}

\newcommand{\sdom}{\prod_{i}\compdom{i}}

\newcommand{\tuple}[1]{\left(#1\right)}

\def\<#1>{\mathinner{\langle#1\rangle}}

\newcommand{\pred}{\pi}
\newcommand{\sysvar}{\mathcal{V}}

\newcommand{\chan}{ch}
\newcommand{\schan}{\rulename{ch}} 

\newcommand{\id}{i}

\newcommand{\cv}{cv}
\newcommand{\scv}{\rulename{cv}} 

\newcommand{\datfun}{{\bf d}}
\newcommand{\sdat}{\rulename{d}} 

\newcommand{\coma}{,\ }

\newcommand{\trans}{\mcal{T}}

\newcommand{\Exp}[1]{2^{#1}}

\newcommand{\msf}[1]{\mathsf{#1}}

\newcommand{\rcp}{{{\rulename{ReCiPe}}}\xspace}
\newcommand{\rcpc}{{{\rulename{R-CHECK}}}\xspace}
\newcommand{\nuxmv}{{{\rulename{nuXmv}}}\xspace}

\newcommand{\keep}{\mbox{\sc keep}}

\newcommand{\typecvar}{{\scriptstyle@\msf{type}}}
\newcommand{\assigncvar}{{\scriptstyle@\msf{asgn}}}
\newcommand{\readycvar}{{\scriptstyle@\msf{rdy}}}

\usepackage{fontawesome}
\usepackage{listings}
  \definecolor{dkgreen}{rgb}{0,0.6,0}
  
\usepackage{color}
\definecolor{dkgreen}{rgb}{0,0.6,0}
\definecolor{gray}{rgb}{0.5,0.5,0.5}
\definecolor{mauve}{rgb}{0.58,0,0.82}
\definecolor{lgray}{gray}{0.9}

\lstdefinelanguage{recipe} {morekeywords={relabel,local,init,repeat,rep,interface,goat,bool,Equals,And,Belong,Not,map,func,loop,if,send,in,comp,receiver,proc,false,true,receive,guard,case,print,NewProcess,Run,Set,function,var,while,int,string,return, SendUpd,GSendUpd,Spawn,Call,WaitUntilTrue,Select,channel}, 
sensitive=false,
aboveskip=3mm,
belowskip=3mm,
showstringspaces=true,
columns=fullflexible,
basicstyle={\sffamily\scriptsize},
morecomment=[l]{//},
morecomment=[s]{/*}{*/},
morestring=[b]", 
mathescape=true,
numbers=left,
numberstyle=\color{black},
keywordstyle=\color{purple},
commentstyle=\color{dkgreen},
stringstyle=\color{mauve},
breaklines=true,
breakatwhitespace=true,
tabsize=2,
numberstyle=\color{black},
rulecolor=\color{black},
xleftmargin={0.75cm},
escapechar=Â£
}

\usepackage{color}
    \definecolor{lightgray}{rgb}{0.95, 0.95, 0.95}
    \definecolor{darkgray}{rgb}{0.5, 0.5, 0.5}
    \definecolor{purple}{rgb}{0.65, 0.12, 0.70}
    \definecolor{blueCode}{rgb}{0, 0, 0.93} 
    \definecolor{greenCode}{rgb}{0, 0.6, 0} 
    \definecolor{redCode}{rgb}{1, 0, 0} 

\lstdefinestyle{rcpStyle}{%
    language=recipe,
    numbers=left,
    basicstyle={\color{blueCode}\sffamily\scriptsize},
    xleftmargin={0.75cm},
    identifierstyle=\color{black},
    keywordstyle=\color{purple},
    keywordstyle = [2]{\color{lime}},
    keywordstyle = [3]{\color{redCode}},
    stringstyle=\color{blueCode}\sffamily,
    commentstyle=\color{greenCode}\sffamily,
    morekeywords = [2]{;,<,>,+,-,!,==,&&,?,:=, <-},
    morekeywords = [3]{(,)},
    tabsize=2,
    showtabs=false,
    showspaces=false,
    showstringspaces=false,
    extendedchars=true,
    breaklines=true,
    escapechar=£
}

\newcommand{\BibTeX}{\rm B\kern-.05em{\sc i\kern-.025em b}\kern-.08em\TeX}
\begin{document}

\title{\rcpc: A Model Checker for Verifying Reconfigurable MAS\thanks{This work is funded by
the ERC consolidator grant
D-SynMA (No. 772459) and the Swedish research council
grants: SynTM (No.
2020-03401) and VR project (No. 2020-04963).}}

\author{
	Yehia Abd
	Alrahman\inst{1}
	\and
	Shaun Azzopardi\inst{1}
	\and
	Nir Piterman\inst{1}
}
\authorrunning{Y. Abd Alrahman et al.}

\institute{University of Gothenburg, Gothenburg, Sweden\\
	\email{\{yehia.abd.alrahman,shaun.azzopardi,nir.piterman\}@gu.se}}
\maketitle 

\begin{abstract}
Reconfigurable multi-agent systems consist of a set of autonomous 
agents, with integrated interaction capabilities that feature opportunistic interaction. Agents seemingly reconfigure their interactions interfaces by forming collectives, and interact based on mutual interests.
 Finding ways to design and analyse the behaviour of these systems is a vigorously pursued research goal. We propose a model checker, named \rcpc\footnote{Find the associated toolkit repository here: \url{https://github.com/dsynma/recipe}.}, to allow reasoning about these systems both from an individual- and a system- level. \rcpc also permits reasoning about interaction protocols and joint missions. \rcpc supports a high-level input language with symbolic semantics, and provides a modelling convenience for interaction features such as reconfiguration, coalition formation, self-organisation, etc.  
 
\end{abstract}

\section{Introduction}\label{sec:intro}
Reconfigurable Multi-agent systems~\cite{HuangCMS16,rmass}, or Reconfigurable MAS for short, emerge as new computational systems, consisting of a set of autonomous agents that interact based on mutual interest, and thus creating a sort of opportunistic interactions. That is, agents seemingly reconfigure their interaction interfaces and dynamically form groups/collectives based on the run-time changes in their execution context. Designing these systems and reasoning about their behaviour is very challenging. This is due to the high-level of dynamism that Reconfigurable MAS exhibit. 

Traditionally, model checking~\cite{mc,LQR17} is considered as a mainstream verification tool for reactive systems~\cite{rs} in the community. A system is usually represented by a low-level language such as NuSMV~\cite{nusmv}, reactive modules~\cite{AH99b,GHW17}, concurrent game structures~\cite{AHK02}, and interpreted
systems~\cite{FHMV95}. The modelling primitives of the latter languages are very close to their underlying semantics, e.g., predicate representation, transition systems, etc. Thus, it makes it hard to model and reason about high-level features of Reconfigurable MAS such as reconfiguration, group formation, self-organisation, etc. Indeed, encoding these features in existing formalisms would not only make it hard to reason about them, but will also create exponentially large and detailed models that are not amenable to verification.
The latter is a classical challenge for model checking and is often termed as \emph{state-space explosion}.

Existing techniques that attempt to tame the state-space explosion
problem (such as BDDs, abstraction, bounded model checking, etc.) can
only act as a mitigation strategy, but cannot provide the right-level
of abstraction to $\qt{\mbox{compactly}}$ model and reason about
high-level features of Reconfigurable MAS.

MAS are often programmed using high-level languages that support domain-specific features of MAS like emergent behaviour~\cite{info19,mmas,scp20}, interactions~\cite{rcp},  intentions~\cite{CohenL90}, knowledge~\cite{FHMV95}, etc. These descriptions are very involved to be directly encoded in plain transition systems. Thus, we often want programming abstractions that focus on the domain concepts, abstract away from low-level details, and consequently reduce the size of the model under consideration.
The rationale is that reasoning about a system requires having the right level of abstraction to represent its behaviour.
Thus, there is a pressing demand to extend traditional model checking tools with support for reasoning about high-level features of Reconfigurable MAS.
This suggests supporting an intuitive description of programs, actions, protocols, reconfiguration, self-organisation, etc.

\rcp~\cite{AbdAlrahmanP21,rcp} is a promising framework to support
modelling and verification of reconfigurable multi-agent system. It is
supported with a symbolic semantics and model representation that
permits the usage of BDDs to enable efficient analysis. However,
writing models in \rcp is very hard and error prone. This is because
\rcp models are encoded in a predicate-based representation that is far
from how we usually program. In fact, the predicate representation of
\rcp supports no programming primitives to control the
structure of programs, and thus everything is encoded using state
variables.

In this paper, we present \rcpc, a model checking toolkit for verifying
and simulating reconfigurable multi-agent systems. \rcpc supports a
minimalistic high-level programming language with symbolic semantics
based on the \rcp framework. We formally present the syntax and
semantics of \rcpc  language and we use it to model and reason about a
nontrivial case study from the realm of reconfigurable and
self-organising MAS. We provide two types of semantics: structural
semantics in terms of automata to recover information about interaction
actions and message exchange, and an execution semantics based on \rcp.
The interaction information recovered in the structural semantics is
recorded succinctly in the execution semantics, and thus permits
reasoning about interaction protocols and message exchange.

We integrate \rcpc with \nuxmv and enable \ltl symbolic and bounded
model checking.
This specialised integration provides a powerful tool that permits
verifying high-level features of Reconfigurable MAS. Indeed, we
can reason about systems both from an individual and a system level. We show how to reason about synchronisations,  interaction protocols, joint missions, and how to express high-level features such as channel mobility, reconfiguration, coalition formation, self-organisation, etc.

The structure of this paper: In Sect. ~\ref{sec:model}, we present a  background on \rcp~\cite{AbdAlrahmanP21,rcp}, the underlying theory of \rcpc. In Sect~\ref{sec:lang}, we present the language of \rcpc and its symbolic semantics.  In Sect. ~\ref{sec:overview}, we provide a nontrivial case study to model autonomous resource allocation. In Sect. ~\ref{sec:logic} we discuss the integration of \rcpc with \nuxmv and we demonstrate our development using high-level properties. Finally, we report concluding remarks in Sect. ~\ref{sec:conc}.
\section{\rcp: a model of computation}\label{sec:model}
We present the underlying theory of \rcpc.
Indeed, \rcpc accepts a high-level language that is based on the symbolic \rcp
formalism \cite{AbdAlrahmanP21,rcp}. We briefly present \rcp agents
and their composition to generate a system-level behaviour.
Formally, agents rely on a set of common variables $\scv$, a set
of data variables $\sdat$, and a set of  channels $\schan$ containing
the broadcast one $\toall$. Common variables $\scv$ are used by agents to  send messages that \qt{indirectly} specify constraints on receivers. That is, each agent has local variables, identified by $\scv$ using a re-labelling function. Thus, agents specify constraints anonymously on common variables which are later translated to the corresponding receiver local variables. That is, when the messages are delivered, the receiver re-label $\scv$ in the constraints and check their satisfaction; data variables $\sdat$ are the actual communicated values in the message; channels $\schan$ define the set of channels that agents use to communicate.

\begin{definition}[Agent]\label{def:comp}
An agent is $
A_{\id}=\langle V_{\id}\coma
f_{\id}\coma \pfunc{g}{s}{\id}\coma
\pfunc{g}{r}{\id},$ $\pfunc{\trans}{s}{\id}\coma \pfunc{\trans}{r}{\id},\theta_{\id}\rangle
$,
\begin{itemize}[label={$\bullet$}, topsep=0pt, itemsep=0pt, leftmargin=10pt]
\item $V_{\id}$ is a finite set of typed local variables, each
  ranging over a finite domain. A state $\cstate{\id}{}$ is an
  interpretation of $V_{\id}$, i.e., if $\mathsf{Dom}(v)$ is the
  domain of $v$, then $\cstate{\id}{}$ is an element in
  $\compdom{\id}$. The set $V'$ denotes the primed copy of $V$ and
$\mathsf{Id}_{\id}$ to denote the assertion $\bigwedge_{v\in
  V_{\id}}v=v'$.

\item $\func{f_{\id}}{\scv}{V_{\id}}$ is a function, associating common variables to local variables.
The notation $f_{\id}$ is used for the assertion $\bigwedge_{\cv\in \scv}\cv=f_{\id}(\cv)$.

\item $\pfunc{g}{s}{\id}(V_{\id}, \schan, \sdat, \scv)$ is a send guard
  specifying a condition on receivers. That is, the predicate, obtained
  from $\pfunc{g}{s}{\id}$ after assigning $\cstate{\id}{}$, $\chan$, and $\datfun$ (an
  assignment to $\sdat$)
  , which is checked against every receiver $j$ after
  applying $f_{j}$.

\item $\pfunc{g}{r}{\id}(V_{\id}, \schan)$ is a receive guard describing
  the connectedness of an agent to a channel $\chan$. We let
  $\pfunc{g}{r}{\id}(V_{\id}, \toall)$ $ = \true$, i.e., every agent
  is always connected to the broadcast channel. Note, however, that
  receiving a broadcast message could have no effect on an agent.
\item $\pfunc{\trans}{s}{\id}(V_{\id}, V'_{\id}, \sdat, \schan)$ is an assertion
describing the send transition relation while $\pfunc{\trans}{r}{\id}(V_{\id}, V'_{\id}, \sdat, \schan)$ is an assertion
describing the receive transition relation.
It is assumed that an agent is broadcast input-enabled, i.e.,
$\forall v, \datfun\ \exists v'\ \such$ $\pfunc{\trans}{r}{\id}(v, v',
\datfun, \toall)$.

\item $\theta_{\id}$ is an assertion on $V_{\id}$ describing the
  initial states, i.e., a state is initial if it satisfies
  $\theta_{\id}$.
\end{itemize}
\end{definition}

Agents exchange messages of the form $m=\tuple{\chan,\datfun,\id,\pred}$. A message is defined
by the channel it is sent on \qt{\chan}, the data it carries \qt{\datfun}, the sender identity \qt{\id},
and the assertion describing the possible local assignments to
common variables of receivers \qt{$\pi$}.
%
The predicate $\pred$ is obtained from
$\pfunc{g}{s}{\id}(\cstate{\id}{},\chan,\datfun,\scv)$ for an agent
$\id$, where $\cstate{\id}{}\in\compdom{\id}$ and $\chan$ and $\datfun$
are the channel and assignment in the observation.

A set of agents agreeing on common variables $\scv$,
data variables $\sdat$, and channels $\schan$ define a \emph{system}.
A system is defined as follows:

\begin{definition}[Discrete System]\label{def:sys}
Given a set $\{A_{\id}\}_{\id}$ of agents, a system is defined as follows: $
S=\conf{\sysvar\coma\rho\coma\theta}
$, where $\sysvar=\bpcup{\id}{V_i}$, a state of the system \qt{$s$} is in
$\sdom$ and
the initial assertion $\theta=\band{\id}{}{\theta_{\id}}$.
The transition relation of the system is as follows:
\[
\begin{array}{l}
\rho:\ \exists \chan\ \exists
\sdat\ \bor{k}{}{\pfunc{\trans}{s}{k}(V_k, V'_k, \sdat, \chan)} \wedge\\
\
\band{j\neq k}{}{}
\tuple{
\exists \scv. f_j\wedge
\tuple{\begin{array}{l r}
\multicolumn{2}{l}{\pfunc{g}{r}{j}(V_j, \chan)\wedge \pfunc{\trans}{r}{j}(V_j, V'_j, \sdat, \chan) \wedge\ \pfunc{g}{s}{k}(V_k, \chan, \sdat, \scv)}
\\[4pt]
\vee &
\neg\pfunc{g}{r}{j}(V_j, \chan)\wedge \mathsf{Id}_j
\\[4pt]
\vee &
\chan=\toall\wedge \neg \pfunc{g}{s}{k}(V_k, \chan, \sdat, \scv)\wedge \mathsf{Id}_j

\end{array}}}
\end{array}
\]
\end{definition}

The transition relation $\rho$ describes two modes of interactions: blocking multicast and non-blocking broadcast. Formally, $\rho$ relates a system state
$\sstate{}$ to its successors $s'$ given a message
 $m=\tuple{\chan,\datfun,k,\pred}$.
Namely, there exists an agent $k$ that sends a message with data $\datfun$
(an assignment to $\sdat$) with assertion $\pred$ (an assignment to
$\pfunc{g}{s}{k}$) on channel $\chan$ and all other agents are either (a)
connected to channel $\chan$, satisfy the send predicate $\pred$, and participate in the
interaction (i.e., has a corresponding receive transition for the message), (b) not connected and idle, or (c) do not satisfy the
send predicate of a broadcast and idle.
That is, the agents satisfying $\pred$ (translated to their local state
by the conjunct $\exists \scv.f_j$) and connected to channel $\chan$
(i.e., $\pfunc{g}{r}{j}(\cstate{j}{}, \chan)$) get the
message and perform a receive transition. As a result of interaction,
the state variables of the sender and these receivers might be
updated.
The agents that are \emph{not connected} to the channel (i.e.,
$\neg\pfunc{g}{r}{j}(\cstate{j}{}, \chan)$) do not
participate in the interaction and stay still.
In case of broadcast, namely when sending
on $\toall$, agents are always connected and the set of receivers
not satisfying $\pred$ (translated again as above) stay still.
Thus, a blocking multicast arises when a sender is blocked until all
\emph{connected} agents satisfy $\pred\wedge f_j$.
The relation ensures that, when sending on
a channel that is different from the broadcast channel $\toall$, the set
of receivers is the full set of \emph{connected} agents.
On the broadcast channel agents who do not satisfy the send
predicate do not block the sender.

%
%
%
%
%
%
%

\rcpc adopts a symbolic model checking approach that directly works on
the predicate representation of \rcp systems. Technically speaking, the
behaviour of each agent is represented by a first-order predicate that
is defined as a disjunction over the send and the receive transition
relations of that agent. Moreover, both send and receive transition
relations can be represented by a disjunctive normal form predicate of
the form $\bor{}{}{(\band{}{}{_j\ \mathsf{assertion}_j})}$. That is, a
disjunct of all possible send/receive transitions enabled in each step
of a computation. In the following, we will define a high-level
language that can be used to write user-friendly programs with symbolic
computation steps. We will also show how to translate these programs to
\rcp predicate representation.

\section{The \rcpc Language}\label{sec:lang}
We formally present the syntax of \rcpc language and show how to
translate it to the \rcp predicate representation. We start by
introducing the type {\ttfamily{agent}}, its structure, and how to
instantiate it; we introduce the syntax of the agent behaviour and how
to create a system of agents. The type {\ttfamily{agent}} is reported
in Fig.~\ref{fig:agent}.
\lstset{language=recipe}
\begin{figure}[h]
\begin{lstlisting}[texcl,mathescape]
    agent name
        local:
            $var\_{name}$:$type$ , $\cdots$, $var\_{name}$:$type$
        init: $\theta_T$
        relabel:
             $common\_var$ <- Exp
	            $\vdots$
	            $common\_var$ <- Exp
	        receive-guard:  $\pfunc{g}{r}{}(V_T, \schan)$
          repeat:  P

\end{lstlisting}
\vspace{-.5cm}
\caption{An agent type}
\label{fig:agent}
\end{figure}

Intuitively, each agent type has a {\ttfamily name} that identifies a
specific type of behaviour. That is, we permit creating multiple
instances/copies with the same type of behaviour. Each agent has a
local state \qt{\ttfamily local} represented by a set of local
variables $V_T$, each of which can be of a type boolean, integer or
enum. The initial state of an agent {\ttfamily init: $\theta_T$} is a
predicate characterising the initial assignments to the agent local
variables. The section \qt{\ttfamily relabel} is used to implement the
relabelling function of common variables in a \rcp agent. Here, we
allow the relabelling to include a binary expression \qt{\ttfamily Exp}
over local variables $V_T$ to accommodate a more expressive relabelling
mechanism, e.g., $\msf{cv}_1\leftarrow(\msf{length}\geq 20)$.
The
section {\ttfamily receive-guard:} $\pfunc{g}{r}{}(V_T, \schan)$
specifies the connectedness of the agent to channels given a current
assignment to its local variables. The non-terminating behaviour of an
agent is represented by {\ttfamily repeat:  P}, which basically
executes the process {\ttfamily P} indefinitely.

Before we introduce the syntax of agent behaviour, we show how to
instantiate an agent and how to compose the different agents to create
a system. An agent type of name $\qt{A}$ can be instantiated as follows
$ A(id,\theta)$. That is, we create an instance of $\qt{A}$ with
identity $id$ and an additional initial restriction $\theta$. Here, we
take the conjunction of $\theta$ with the predicate in the {\ttfamily
init} section of the type $\qt{A}$ as the initial condition of this
instance. We use the parallel composition operator $\|$ to inductively
define a system as in the following production rule:
\[\textrm{(System)}  \qquad S ::=\quad  A(id,\theta) \quad |\quad S_1\|S_2
\]

That is, a system is either an instance of agent type or a parallel composition of set of instances of (possibly) different types. The semantics of $\|$ is fully captured by $\rho$ in Def.~\ref{def:sys}.

The syntax of an \rcpc process is inductively defined as:
\begin{table}[h]
\centering
\begin{tabular}{ll}
%
%
\textrm{(Process)}   & $P ::=\quad   P ; P\quad |\quad P + P\quad |\quad {\mathsf{rep\ } P}\quad |\quad C $\\

\medskip

\textrm{(Command)} & $C ::=\quad  l:C \quad |\quad\langle\Phi\rangle\ \chan\ !\ \pred\ \datfun\ \mathsf{U} \quad |\quad  \langle\Phi\rangle\ \chan\ ?\ \mathsf{U}$\\
\end{tabular}
\label{tab:syntax}
\end{table}

A process $P$ is either a sequential composition of two processes $P;P$, a non-deterministic choice between two processes $P+P$, a loop $\mathsf{rep}\ P$, or a command $C$. There are three types of commands corresponding to either a labelled command, a message-send or a message-receive. A command of the form $l:C$ is a syntactic labelling and is used to allow the model checker to reason about syntactic elements as we will see later. A command of the form $\langle\Phi\rangle\ \chan\ !\ \pred\ \datfun\ \mathsf{U}$ corresponds to a message-send.  Intuitively, the predicate $\Phi$ is an assertion over the current assignments to local variables, i.e., is a pre-condition that must hold before the transition can be taken. As the names suggest, $\chan$, $\pred$ and (respectively) $\datfun$ are  the communication channel, the sender predicate, and the assignment to data variables (i.e., the actual content of the message). Lastly, $\mathsf{U}$ is the next assignment to local variables after taking the transition. We use \qt{!} to distinguish send transitions. A command of the form $\langle\Phi\rangle\ \chan\ ?\ \mathsf{U}$ corresponds to a message-receive. Differently from message-send, the predicate $\Phi$ can also predicate on the received values from the incoming message, i.e., the assignment $\datfun$. We use \qt{?} to distinguish receive transitions.

Despite the minimalistic syntax of \rcpc, we can express every control
flow structure in a high-level programming language. For instance, by
combining non-determinism and pre-conditions of commands, we can encode
any structure of {IF-Statement}. Similarly, we can encode finite loops
by combining $\mathsf{rep}\ P$ and commands $C$, e.g., $(\mathsf{rep}\
C1 + C_2)$ means: repeat $C_1$ or block until $C_2$ happens.

\subsection{The semantics of \rcpc}\label{sec:trans}
We initially give a structural semantics to \rcpc process using a finite automaton such that each transition in the automaton corresponds to a symbolic transition.
Intuitively, the automaton represents the control structure of an \rcpc process. We will further use this automaton alongside the agent definition to give an \rcpc agent an execution semantics based on the symbolic \rcp framework. This two-step semantics will help us in verifying structural properties about \rcpc agents.

\begin{definition}[Structure automaton]\label{def:struct} A structure automaton is of the form $G=\conf{S,\ \Sigma,\ s_i,\ E,\ s_f}$, where
\begin{itemize}
\item $S$ is a finite set of states;

\item $s_i,s_f\in S$: are two states that, respectively, represent the \emph{initial} state and the \emph{final} state in $G$;

\item $\Sigma$ is the alphabet of $G$;

\item $E\subseteq S\times \Sigma\times S$: is the set of edges of $G$.
\end{itemize}

We use $(s_1,\sigma,s_2)$ to denote an edge $e\in E$ such that $s_1$ is
the source state of $e$,  $s_2$ is the target state of $e$ and the
letter $\sigma$ is the label of $e$.
\end{definition}

Now, everything is in place to define the structure semantics of \rcpc processes. We define a function $\llparenthesis\ \centerdot\ \rrparenthesis^{[s_i,s_f]}: P \rightarrow \Exp{E}$ which takes an \rcpc process $P$ as input and produces the set of edges of the corresponding structure automaton. The function $\llparenthesis\ \centerdot\ \rrparenthesis^{[s_i,s_f]}$ assumes that each process has unique initial state $s_i$ and final state $s_f$ in the structure automaton. Please note that the states of the structure automaton only represent the control structure of the process, and an agent can have multiple initial states depending on $\theta_T$ while starting from $s_i$. The definition of the translation function $\llparenthesis\ \centerdot\ \rrparenthesis^{[s_i,s_f]}$ is reported below:
\begin{table}[h]
\centering
\begin{tabular}{ccc}

$
{
\begin{array}{@{}l@{\ }r@{\,\,}c@{\,\,}l@{}}
\llparenthesis \mathsf{repeat:}\ P\rrparenthesis^{[s_i,s_f]}  \triangleq  \llparenthesis P\rrparenthesis^{[s_i,s_i]}
 \\ [2ex]
\llparenthesis P_1;P_2\rrparenthesis^{[s_i,s_f]}  \triangleq  \llparenthesis P_1\rrparenthesis^{[s_i,s_1]} \bcup{}{}{} \llparenthesis P_2\rrparenthesis^{[s_1,s_f]}
\qquad\mbox{for a fresh $s_1$}

 \\ [2ex]

\llparenthesis P_1+P_2\rrparenthesis^{[s_i,s_f]}  \triangleq  \llparenthesis P_1\rrparenthesis^{[s_i,s_f]} \bcup{}{}{} \llparenthesis P_2\rrparenthesis^{[s_i,s_f]}
 \\ [2ex]

\llparenthesis \mathsf{rep}\ P\rrparenthesis^{[s_i,s_f]}  \triangleq  \llparenthesis P\rrparenthesis^{[s_i,s_i]}
 \\ [2ex]

 \llparenthesis C\rrparenthesis^{[s_i,s_f]}  \triangleq  \set{(s_i,C,s_f)}
\end{array}}
 $
 \end{tabular}
\label{tab:encoding}
\end{table}

Intuitively, the structure semantics of $\llparenthesis \mathsf{repeat:}\ P\rrparenthesis^{[s_i,s_f]}$ corresponds to a self-loop in the structure automaton (with $s_i$ as both the source and the target state ) and where $P$ is repeated indefinitely. Moreover, the semantics $\llparenthesis P_1;P_2\rrparenthesis^{[s_i,s_f]}$ is the union of the transitions created by $P_1$ and $P_2$ while creating a fresh state in the graph $s_1$ to allow sequentiality, where $P_1$ starts in $s_i$ and ends in $s_1$ and later $P_2$ continues from $s_1$ and ends in $s_f$. That is, the structure of the process is encoded using an extra memory. Differently, the non-deterministic choice $\llparenthesis P_1+P_2\rrparenthesis^{[s_i,s_f]}$ does not require extra memory because the execution  $P_1$ and $P_2$ is independent. The semantics of $\llparenthesis \mathsf{rep}\ P\rrparenthesis^{[s_i,s_f]}$ is similar to $\llparenthesis \mathsf{repeat:}\ P\rrparenthesis^{[s_i,s_f]}$ and is introduced to allow self-looping inside a non-terminating process. Finally, the semantics of a command $C$ in an \rcpc process corresponds to an edge $\set{(s_i,C,s_f)}$ in the structure automaton. This means that the alphabet $\Sigma$ of the automaton ranges over \rcpc commands. Note that the translation function is completely syntactic and does not involve evaluation or enumeration of variables, and thus the resulting automaton is symbolic.

To translate an \rcpc agent into a \rcp agent, we first introduce the
following  functions:$ \qt{\mathsf{typeOf}}$, $\qt{\mathsf{varsOf}}$,
$\qt{\mathsf{predOf}}$ and $\qt{\mathsf{guardOf}}$ on a command $C$.
That is, $\qt{\mathsf{typeOf}(C)}$ returns the type of a command $C$ as
either $!$ (send) or $?$ (receive).  For example, the
$\mathsf{typeOf}(\langle\Phi\rangle\ \chan\ !\ \pred\ \datfun\
\mathsf{U})=\ !$; the $\qt{\mathsf{varsOf}(C)}$ function returns the
set of local variables that are updated in $C$; the function
$\qt{\mathsf{predOf}(C)}$ returns the equivalent predicate
characterising $C$ (while excluding the send predicate $\pred$ in send
commands). For instance, $\qt{\mathsf{predOf}(\msf{\langle
Link=c\rangle \toall! \pi(MSG := m)[Link := b]})}$ is the predicate
$\qt{\msf{(Link=c)\wedge (\chan=\toall)\wedge (MSG = m)\wedge (Link' =
b)}}$. That is, the predicate characterising local variables $V_T$, the
primed copy $V'_T$, the channel $\chan$ and the data variables $\sdat$;
and finally $\qt{\mathsf{guardOf}(C)}$ returns the send predicate
$\pred$ in a send command and false otherwise.

Moreover, we use $\keep(X)$ to denote that the set of local variables $X$ is
  not changed by a transition (either send or receive).
  More precisely, $\keep(X)$ is equivalent to the following assertion
  $\bigwedge_{x\in X}x=x'$ where $x'$ is the primed copy of $x$.

The following definition shows how to construct a \rcp agent from an \rcpc agent with structure semantics interpreted as a structure automaton.

\begin{definition}[from \rcpc to \rcp]\label{def:trans} Given an instance of agent type $T$ as defined in Fig.~\ref{fig:agent} with a structure semantics interpreted as a structure automaton $G=\conf{S,\ \Sigma,\ s_i,\ E,\ s_f}$, we can construct a \rcp agent $
A=\langle V\coma
f\coma \pfunc{g}{s}{}\coma
\pfunc{g}{r}{},$ $\pfunc{\trans}{s}{}\coma \pfunc{\trans}{r}{},\theta\rangle
$ that implements its behaviour.

We construct $A$ as follows:
\begin{itemize}
\item $V=V_T \cup \set{\mathsf{st}}$: that is, the union of the set of declared variables $V_T$ in the {\ttfamily local} section of $T$ in Fig.~\ref{fig:agent} and a new state variable $\qt{\mathsf{st}}$ ranging over the states $S$ in $G$ of the structure automaton, representing the control structure of the process of $T$. Namely, the control structure of the behaviour of $T$ is now encoded as an additional variable in $A$;

\item the initial condition $\theta=\theta_T\wedge(\mathsf{st}=s_i)$: that is the conjunction of  the initial condition $\theta_T$ in the {\ttfamily init} section of $T$ in Fig.~\ref{fig:agent} and the predicate $\mathsf{st}=s_i$, specifying the initial state of $G$.

\item $f$ and $g^r$ have one-to-one correspondence in section  {\ttfamily relabel}  and  section {\ttfamily receive-guard} respectively of $T$ in Fig.~\ref{fig:agent}.

\item $g^s=\bor{\sigma\in \Sigma:\ \mathsf{typeOf}(\sigma)=\ !}{}{\mathsf{guardOf}(\sigma)}$

\item $\pfunc{\trans}{s}{}=$
\[\bor{(s_1,\sigma,s_2)\in E:\ \mathsf{typeOf}(\sigma)=\ !}{}{\tuple{\begin{array}{c}
\mathsf{predOf}(\sigma)\wedge (\mathsf{st}=s_1)\wedge(\mathsf{st}'=s_2)\wedge\\
 \keep{(V_T\backslash\mathsf{varsOf}(\sigma))}
\end{array}}}\]

\item $\pfunc{\trans}{r}{}=$
\[\bor{(s_1,\sigma,s_2)\in E:\ \mathsf{typeOf}(\sigma)=\ ?}{}{\tuple{\begin{array}{c}
\mathsf{predOf}(\sigma)\wedge (\mathsf{st}=s_1)\wedge(\mathsf{st}'=s_2)\wedge\\
 \keep{(V_T\backslash\mathsf{varsOf}(\sigma))}
\end{array}}}\]
\end{itemize}
\end{definition}

\lstset{language=recipe, style=rcpStyle}

\section{autonomous resource allocation}\label{sec:overview}
We model a scenario where a group of clients are requested to jointly solve a problem. Each client will buy a computing virtual machine (VM) from a resource manager and use it to solve its task. Initially, clients know the communication link of the manager, but they need to self-organise and coordinate the use of the link anonymously. The manager will help establishing connections between the clients and the available machines, and later clients proceed interacting independently with machines on private links that they learn when the connection is established.

There are two types of machines: high performance machines and standard ones.
The resource manager commits to provide high performance VMs to
clients, but when all of these machines are reserved, the clients are
assigned to standard ones. The protocol proceeds until each client buys a machine, and then  all clients have to collaborate to solve the problem and complete the task.

A client uses the local variables
$\qt{\msf{cLink,\ mLink,\ tLink,\ role}}$ to control its behaviour, where
$\qt{\msf{cLink}}$ stores a common link (i.e., the link to interact with the resource manager),
$\qt{\msf{mLink}}$ is a placeholder for a mobile link that can be learnt at run-time, $\qt{\msf{tLink}}$ is a link to synchronise with other clients to complete the task,
and $\qt{\msf{role}}$ is the role of the client.
The initial condition $\msf{\theta_c}$ of a client is:
\[
\begin{array}{rcl}
\msf{\theta_{c}} : \msf{cLink=c}\ \msf{\wedge\ mLink=\msf{empty}\wedge \ tLink=\msf{t}\wedge role=client,}
\end{array}
\]
specifying that the resource
manager is reachable on $\msf{c}$, the mobile link is empty
, the task link is $\qt{\msf{t}}$ and the role is $\msf{client}$.

Note that the interfaces of agents are parameterised to their
local states and state changes may create dynamic and opportunistic interactions. For instance, when $\msf{cLink}$ is set to $\msf{empty}$, the client
discards all messages on $\msf{c}$; also when a run-time channel is
assigned to $\msf{mLink}$, the client starts receiving messages on that channel.

Clients may use broadcast or multicast;
in a broadcast, receivers (if exist) may anonymously receive the
message when they are interested in its values (and when they satisfy
the send predicate). Otherwise, an agent may not participate in the
interaction.
In multicast, all agents listening on the multicast channel must
participate to enable the interaction.

Broadcast is used when agents are unaware of
the existence of each other while (possibly) sharing some resources
while multicast is used to capture a more structured interaction where
agents have dedicated links to interact. In our example, clients
are not aware of the existence of each other while they share the
resource manager channel $\msf{c}$. Thus they may coordinate
to use the channel anonymously by means of broadcast.
A client reserves the channel $\msf{c}$ by means of a broadcast
message with a predicate targeting agents
with a client role. All other clients self-organise and disconnect from $\msf{c}$
and wait for a release message.

A message in \rcpc carries an assignment to a set of data
variables $\sdat$. In our scenario, $\sdat=\set{\rulename{msg}, \rulename{lnk}}$
where \rulename{msg} denotes the label of the message and takes
values from: \[{\msf{reserve}, \msf{request}, \msf{release}, \msf{buy}, \msf{connect},
\msf{full}, \msf{complete}}\]
Moreover, \rulename{lnk} is used to exchange a  link with
other agents.

Agents in this scenario use one common variable $\msf{cv}$ ranging over
roles to specify potential receivers. Remember that every agent $i$ has
a relabelling function $f_i:\scv\rightarrow V_i$
that is applied to the send guard once a message is
delivered to check whether it is eligible to receive.
For a client, $f_c(\msf{cv})=\msf{role}$.
The send guard of a client appears in the messages that
the client sends, and we will explain later. In general,
broadcasts are destined to agents assigning to the common variable
$\msf{cv}$ a value matching the
 $\msf{role}$ of the sender, i.e, $\msf{client}$; messages
on $\msf{cLink}$ are destined to agents assigning $\msf{mgr}$ to $\msf{cv}$;
and other messages are destined to everyone listening on the right channel.

The receive guard is:
$
\begin{array}{rcl}
\pfunc{g}{r}{c} : (\msf{\chan=\star) \vee (\chan=cLink) \vee(\chan=tLink)}.
\end{array}
$
That is, reception is always enabled on broadcast and on a channel that matches
the value of  $\msf{cLink}$ or $\msf{tLink}$.
Note that these guards are parameterised to local variables and
thus may change at run-time, creating a dynamic communication structure.

The behaviour of the client is reported in Fig.~\ref{fig:client} below:
\begin{figure}[h]\small
\begin{lstlisting}[texcl,mathescape]
repeat: (
            (sReserve: <cLink==c > *! (cv==role)(MSG := reserve)[]
            +
            rReserve: <cLink==c && MSG == reserve> *? [cLink := empty]
            )
	    	    ;
            (
            sRequest: <cLink!=empty> cLink! (cv==mgr)(MSG := request)[];

             rConnect: <mLink==empty && MSG == connect> cLink? [mLink := LNK];

             sRelease: <TRUE> *! (cv==role)(MSG := release)[cLink := empty];

             sBuy: <mLink!=empty> mLink! (TRUE)(MSG := buy)
             [mLink := empty];
             (
                sSolve: <TRUE> tLink!(TRUE)(MSG := complete)[]
               +
                <MSG == complete> tLink? []
             )
            +
             rRelease: <cLink==empty && MSG == release> *? [cLink := c]
            )
          )
\end{lstlisting}
\vspace{-.5cm}
\caption{Client Behaviour}
\label{fig:client}
\end{figure}

In this example, we label each command with a name identifying the message and its type (i.e., \qt{$\msf{s}$} for send and \qt{$\msf{r}$} for receive). For instance, the send transition at Line $2$ is labelled with \qt{$\msf{sReserve}$} while the receive transition at Line $4$ is labelled with \qt{$\msf{rReserve}$}. We will use them later to reason about agent interactions syntactically.

Initially in Lines $2-5$, every client may either broadcast a \qt{$\msf{reserve}$} message
to all other clients (i.e., $\msf{(cv=role)}$) or receive a \qt{$\msf{reserve}$} message from one of them. This is to allow clients to self-organise and coordinate to use the common link.
That is, a client may initially reserve an interaction session with the resource
manager by broadcasting a \qt{$\msf{reserve}$} message to
 all other clients, asking them to disconnect the common link $c$ (stored in
 their local
 variable $\msf{cLink}$); or receive a \qt{$\msf{reserve}$} message,
  i.e., gets asked by another client to disconnect channel $c$.
  In either case, the client progress to Line $7$. Depending on what happened in the previous step, the client may proceed to establish a session with the resource manager (i.e., $\msf{(cv=\msf{mgr})}$) and a machine (Lines $8-20$) or gets stuck waiting for a \qt{$\msf{release}$} message from the client, currently holding the session (Line $22$). In the latter case, the client gets back in the loop to (Line $1$) after receiving a \qt{$\msf{release}$} message and attempts again to establish the session.

 In the former case, the client uses the blocking multicast channel $\msf{c}$ to send a request to the resource
 manager (Line $8$) and waits to receive a private connection link with a virtual machine agent on \qt{$\msf{cLink}$} (Line $10$). When the client receives the \qt{$\msf{connect}$} message on $\msf{cLink}$, the client assigns its $\msf{mLink}$ variable the value of $\rulename{lnk}$ in the message. That is, the client is now ready to communicate on $\msf{mLink}$. On Line $12$, the agent releases the common link $\msf{c}$ by broadcasting a release message to all other clients (with $\msf{(cv=role)}$) and proceeds to Line $14$ and starts communicating privately with the assigned VM agent. The client buys a service from the VM agent on a dedicated link
 stored in $\msf{mLink}$ by sending a \qt{$\msf{buy}$} to the VM agent to complete the transaction. The client proceeds to line $16$ and wait for other clients to collaborate and finish the task. Thus, the client either initiates the last step and sends a $\msf{complete}$ message when the rest of clients are ready (Line $17$) or receives a $\msf{complete}$ message from one of them when the client is ready.

We may now specify the manager and the virtual machine behaviour, and
show how reconfigurable multicast can be used to model a point-to-point interaction in
a clean way.

The resource manager has the following local variables:
\[{\msf{hLink,\ sLink,\ cLink,\ role}}\]
where $\msf{hLink}$ and $\msf{sLink}$ store channel names to
communicate with high- and standard-performance VMs respectively and
the rest are as defined before.

The initial condition is:
\[
\msf{\theta_{m}}: \msf{hLink=g_1\wedge sLink=g_2\wedge cLink=c\wedge role=mgr}
\]

Note that the link $\msf{g_1}$ is used to communicate with the group of high performance machines while $\msf{g_2}$ is used for standard ones.

The send guard for a manager is always satisfied, (i.e.,
$\pfunc{g}{s}{m}$ is  $\msf{ \true}$) while the
receive guard specifies that a manager only receives broadcasts or on
channels that match with values of
$\msf{cLink}$ or $\msf{hLink}$ variables, i.e., $\pfunc{g}{r}{m}$ is
$\msf{(\chan=\star) \vee (\chan=cLink) \vee (\chan=hLink)}$.

The behaviour of the agent manager is reported in Fig.~\ref{fig:manager} below:

\begin{figure}[h]\small
\begin{lstlisting}[texcl,mathescape]
 repeat: (
            rRequest: <MSG == request> cLink? [];
            sForward: <TRUE> hLink! (TRUE)(MSG := request)[];
            (
             rConnect: <MSG == connect> cLink? []
             +

             rep ( rFull: <MSG == full> hLink? [];
                     sRequest: <TRUE> sLink! (TRUE)(MSG := request)[]
                 )
            )
            )
\end{lstlisting}
\vspace{-.5cm}
\caption{Manager Behaviour}
\label{fig:manager}
\end{figure}

In summary, the manager initially forwards requests received on channel $\msf{c}$ (Line $2$) to the high performance
 VMs first as in (Line $3$). The negotiation protocol with machines is reported in Lines $5-10$. The manager can receive a \qt{$\msf{connect}$} message and directly enable the client to connect with the virtual machine as in (Line $5$) or receive a \qt{$\msf{full}$} message, informing that all high performance machines are fully occupied. In the latter case, the requests are forwarded to the standard performance machines on $\msf{sLink}$ as in (Lines $8-10$). The process repeats until a \qt{$\msf{connect}$} message is received (Line $5$) and the manager gets back to (Line $1$) to handle other requests.
Clearly, the specifications of the manager assumes that there are a plenty of standard VMs and
a limited number of high performance ones. Thus it only expects a $\msf{full}$ message to be
received on channel $\msf{hLink}$. Note also that the manager gets ready to handle the next request
once a connect message ($\msf{connect}$) is received on channel $\msf{c}$ and leaves the client
and the selected VM to interact independently.

The virtual machine has the following local variables: \[{\msf{gLink,\ pLink,\  cLink,\ asgn}}\] where
\qt{$\msf{asgn}$} indicates if the VM is assigned, \qt{$\msf{gLink}$} is a group link, \qt{$\msf{pLink}$} is a private link
and the rest is as before; apart from \qt{$\msf{gLink}$} and \qt{$\msf{pLink}$},
which are machine dependent, the initial condition is of the form:
\[
\begin{array}{rcl}
\msf{\theta_{vm}}: \msf{\neg asgn \wedge cLink=empty}
\end{array}
\]
 where initially virtual machines are not listening on the common link $\msf{cLink}$. Depending on the group that the machine belong to, the \qt{$\msf{gLink}$} will either be assigned to high performance machine group {$\qt{\msf{g_1}}$} or the standard one $\qt{\msf{g_2}}$. Moreover, each machine has a unique private link $\qt{\msf{pLink}}$.
The send guard for a VM is always satisfied, (i.e., $\pfunc{g}{s}{\msf{mv}}$ is  $\msf{ \true}$) while the
receive guard specifies that a VM always receives on broadcast, {$\msf{pLink}$}, $\msf{gLink}$ and  $\msf{cLink}$, i.e.,
\[
\begin{array}{c}
\pfunc{g}{r}{\msf{vm}} : \msf{\chan=\toall\ \vee\  \chan=gLink\ \vee}
\ \msf{\chan=pLink\vee\ \chan=cLink}
\end{array}
\]

The behaviour of the virtual machine agent is reported in Fig.~\ref{fig:machine}:
\begin{figure}[h]\small
\begin{lstlisting}[texcl,mathescape]
  repeat: (
            rForward: <cLink==empty && MSG == request> gLink? [cLink:= c];
            (
             sConnect: <cLink==c && !asgn> cLink! (TRUE)(MSG := connect, LNK := pLink)[cLink:= empty, asgn:= TRUE]
             +
             sFull: <cLink==c && asgn> gLink! (TRUE)(MSG := full)[cLink:= empty]
             +
             rConnect: <cLink==c && MSG == connect> cLink? [cLink:= empty]
             +
             rFull: <cLink==c && asgn && MSG == full> gLink? [cLink:= empty]
            )
            +
            rBuy: <MSG == buy> pLink? []
            )
\end{lstlisting}
\vspace{-.5cm}
\caption{Machine Behaviour}
\label{fig:machine}
\end{figure}

Intuitively, a VM either receives the forwarded request on the group channel $\msf{gLink}$ (Line $2$)
and thus activating the common link and also a
nondeterministic choice between $\msf{connect}$ and $\msf{full}$
messages (Lines $3-11$) or receives a $\msf{buy}$ message from a client on the private link $\msf{pLink}$.
In the latter case, the VM agent agrees to sell the service and stays idle.
In the former case,  a VM sends $\msf{connect}$ with its private
link $\msf{pLink}$ carried on the data variable $\rulename{lnk}$ and send it
on $\msf{cLink}$ if it is not
assigned or sends $\msf{full}$ on $\msf{gLink}$ otherwise. Note
that $\msf{full}$ message can only go through if all VMs in group
$\msf{gLink}$ are assigned. Note that reception on $\msf{gLink}$ is always enabled by the receive guard $\pfunc{g}{r}{\msf{vm}}$ and the receive transition at Line $10$ specifies that a machine enables a send on a $\msf{full}$ message only when it is assigned. For example, if $\msf{gLink=g_1}$ then only when all machines in group $g_1$ are assigned, a $\msf{full}$ message can be enabled.

Furthermore, a $\msf{connect}$
 message will also be
received by other VMs in the group
$\msf{gLink}$ (Line $8$). As a result, all other available VMs (i.e., $\msf{\neg
  asgn}$) in the same group do not reply to the request. Thus, one VM
is non-deterministically selected to provide a service and a
point-to-point like interaction is achieved.
Note that this easy encoding is possible because agents change
communication interfaces dynamically by enabling and disabling 
channels.

Now, we can easily create an \rcpc system as  follows:
\begin{equation}\label{eq:sys}
\begin{array}{cc}
\msf{system}  = \msf{Client(client1,TRUE)}\ \|\ \msf{Client(client2,TRUE)} \\ \qquad\|\  \msf{Client(client3,TRUE)}\ \|\ \msf{Manager(manager,TRUE)}\\ \qquad\|\ \msf{Machine(machine1,gLink=g1  \wedge pLink=vmm1)} \\ \qquad\|\ \msf{Machine(machine2,gLink=g1  \wedge pLink=vmm2)} \\ \qquad\|\ \msf{Machine(machine3,gLink=g2  \wedge pLink=vmm3)}
\end{array}
\end{equation}

This system is the parallel composition (according to Def.~\ref{def:sys}) of three copies of a client $\set{\msf{client_1},\dots,\msf{client_3}}$; a one copy of a manager {$\msf{manager}$}; and finally three copies of a machine $\{\msf{machine_1},\dots,$ $\msf{machine_3}\}$, each belongs to a specific group and a private link. For instance,  $\msf{machine_1}$ belongs to group $\qt{\msf{g_1}}$ (the high performance machines) and has a private link named $\qt{\msf{vmm1}}$.
\section{\nuxmv and Model-checking}\label{sec:logic}
We describe the integration of \rcpc with the \nuxmv model checker~\cite{CimattiG12} to enable an enhanced symbolic \ltl model-checking.
We also demonstrate our developments using examples. We will show how the combined features of \rcpc, the symbolic \ltl model-checking, and \nuxmv provides a powerful tool to verify high-level features of reconfigurable and interactive systems.

\paragraph{\bf From \rcpc to \nuxmv} We give individual \rcpc agents a symbolic semantics based on the \rcp framework as shown in Sect.~\ref{sec:trans} and Def.~\ref{def:trans}. Notably, we preserve the labels of commands (i.e., $l:\sigma$) and use them as subpredicate definitions. For instance, given a labeled edge $(s_1,l:\sigma,s_2)$ in the structure automaton $G$ in Def.~\ref{def:struct}, we translate it into the following predicate in \rcp as explained in Def.~\ref{def:trans}: \[{
l :=\mathsf{predOf}(\sigma)\wedge (\mathsf{st}=s_1)\wedge(\mathsf{st}'=s_2)\wedge
 \keep{(V_T\backslash\mathsf{varsOf}(\sigma))}
}\]

The only difference here is that the label $\qt{l}$ is now a predicate definition and its truth value defines if the transition $(s_1,l:\sigma,s_2)$ is feasible. Since every command is translated to either message-send or message-receive, we can use these labels now to refer to message exchange syntactically inside \ltl formulas.

Moreover, we rename all local variables of agents to consider the
identity of the agent as follows: for example, given the
\qt{$\msf{cLink}$} variable of a client, we generate the variable
\qt{$\msf{client-cLink}$}. This is important when different agents use
the same identifier for local variables. We also treat all data
variables $\sdat$ and channel names $\schan$ as constants and we
construct a \rcp system $S=\conf{\sysvar\coma\rho\coma\theta}$ as
defined in Def.~\ref{def:sys} while considering subpredicate
definitions and agent variables after renaming. Technically, a \rcp
system $S$ has a one-to-one correspondence to a \nuxmv module $M$. That
is, both $S$ and $M$ agrees on local variables $\sysvar$ and the
initial condition $\theta$, but are slightly different with respect to
transition relations. Indeed, the transition relation $\rho$ of $S$ as
defined in Def.~\ref{def:sys} is translated to an equivalent transition
relation $\hat{\rho}$ of $M$ as follows:
\[\hat{\rho}=\ \rho \lor\ (\neg\rho\wedge\keep(\sysvar))\]
That is, \nuxmv translates deadlock states in $S$ into stuttering (sink) states in $M$ where system variables do not change.

\begin{figure}
\centering
\begin{tabular}{c}
$
\begin{array}{c}
\includegraphics[scale=.25]{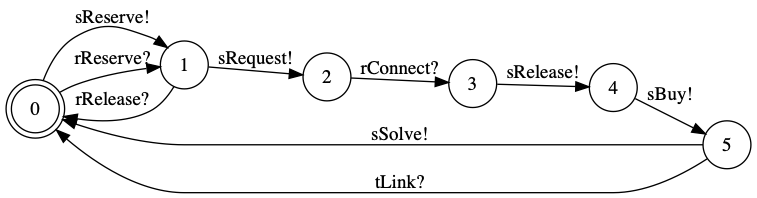}\\
\mbox{(a) Client}\\
\end{array}$ \\
$
\begin{array}{cc}
\begin{array}{c}
\includegraphics[scale=.15]{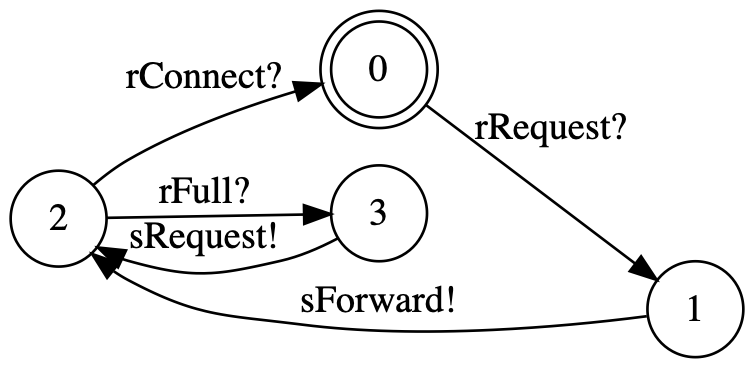}  \\
\mbox{(b) Manager}\\
\end{array} &
\begin{array}{c}
\includegraphics[scale=.15]{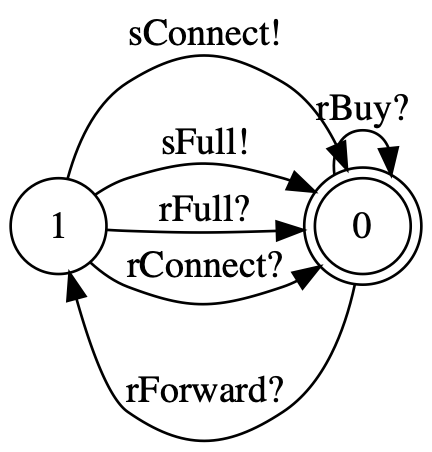}\\
\mbox{(c) Machine}\\
\end{array}
\end{array}$
\end{tabular}
\vspace{-.4cm}
\caption{symbolic structure automata}
\label{fig:sauto}
\end{figure}

\rcpc provides an interactive simulator that allows the user to
simulate the system either randomly or based on predicates that the
user supplies. For instance, starting from some state in the
simulation, the user may supply the constraint
$\msf{next(client1{-}cLink)=c}$ to ask the simulator to select the
transition that leads to a state where the next value of
$\msf{client1{-}cLink}$ equals \qt{$\msf{c}$}. If such constraint is
feasible (i.e., there exists a transition satisfying the constraint),
the simulator selects such transition, and otherwise it returns an
error message. Users can also refer to message -send and -receive using
command labels in the same way. A constraint on a send transition like
\qt{$\msf{client1{-}sReserve}$}, to denote the sending of the message
\qt{$\msf{reserve}$} in Fig.~\ref{fig:client}, Line $2$, means that
this transition is feasible in the current state of simulation.
However, a constraint on a receive transition
\qt{$\msf{client{-}rReserve}$}, like on the message in
Fig.~\ref{fig:client}, Line $4$, means that this transition is already
taken from the previous state of simulation. This slight difference
between send and receive transitions is due to the fact that receive
transition cannot happen independently and only happen due to a joint
send transition. Finally, \rcpc is supported with an editor, syntax
highlighting and visualising tool. For instance, once the model of the
scenario in Sect.~\ref{sec:overview} is complied, \rcpc produces the
corresponding labelled and symbolic structure automata in
Fig.~\ref{fig:sauto}, and thus the user may use these automata to
reason about interactions.

\paragraph{\bf Symbolic Model Checking}
\rcpc supports both symbolic \ltl model checking and bounded \ltl model
checking. We illustrate the
 capabilities of \rcpc by several examples. In the rest of
the section, we will use Equation~\ref{eq:sys},
Sect.~\ref{sec:overview} and the corresponding structure automata in
Fig.~\ref{fig:sauto} as the system under consideration. 

We show how to verify properties about agents both from individual and interaction protocols level by predicating on message exchange rather than on atomic propositions.
It should be noted that the transition labels in Fig.~\ref{fig:sauto} are not mere labels, but rather predicates with truth values changing dynamically at run-time, introducing opportunistic interaction.
For instance, we can reason about a client and its connection to the
system as follows:
\[\begin{array}{cc}
\msf{G\ (client1{-}sReserve \rTo{} F\ client1{-}sRequest)}&(1)\\
\msf{G\ (client1{-}sReserve \rTo{} F\ client1{-}sRelease)}&(2)\\
\msf{G\ (client1{-}sRequest \rTo{} F\ client1{-}rConnect)}&(3)
\end{array}\]

The liveness condition $(1)$ specifies that the client can send a request
to the manager after it has already reserved the common link
\qt{$\msf{c}$}; the liveness condition $(2)$ specifies that the client
does not hold a live lock on the common link \qt{$\msf{c}$}. Namely,
the client releases the common link eventually.  The liveness condition
$(3)$ specifies that the \emph{system} is responsive, i.e., after the
client's request, other agents collaborate to eventually supply a
connection.

We can also reason about synchronisation and reconfiguration in relation to local state as in the following:
\[\begin{array}{cc}
\msf{G\ (manager{-}sForward \rTo{} X\ machine1{-}rForward)}&(4)\\
\msf{F\ (client1{-}sRelease\ \&\ G(!client1{-}rConnect))}&(5)\\
\msf{G\ ((!machine1{-}asgn\ \&\ machine1{-}rForward)}\\ \rTo{}
\msf{machine1{-}sConnect)}&(6)
\end{array}\]

In $(4)$, we refer to synchronisation, i.e., the manager has to forward the request before the  machine can receive it. Note that this formula does not hold for $\msf{machine_3}$ because $\msf{sForward}$ is destined for group $\msf{g_1}$; we can refer to reconfiguration in $(5)$, i.e., eventually the client disconnects from the common link \qt{$\msf{c}$}, and it can never be able to receive connection on that link; moreover, in $(6)$ the machine sends a connection predicated on its local state, i.e., if it is not assigned. Note that $(6)$ does not hold because $\msf{machine_1}$ might lose the race for $\msf{machine_2}$ in group $\msf{g_1}$ to execute $\msf{connect}$ message.

We can also specify channel mobility and joint missions from a declarative and centralised point of view.
\[\tuple{
\begin{array}{lc}
\msf{F(client1{-}mLink\neq empty)\ \&}\\
\qquad\qquad\msf{F\ (client2{-}mLink\neq empty)\  \&}\\
\hfill \msf{F\ (client3{-}mLink\neq empty))}\rTo{\ \ \ \ }\\ \msf{ F\
(client1{-}sSolve\ |\ client2{-}sSolve\ |\ client3{-}sSolve)}
\end{array}
}\]

That is, every client will eventually receives a mobile link (i.e., its $\msf{mLink\neq empty}$) where it will use this private link to buy a VM, and eventually one client will initiate the termination of the mission by synchronising with the other clients to solve the joint problem.

We are unaware of a model-checker that enables reasoning at such a high-level. The full tool support and all examples in this paper are attached to the submission as supplementary material.

\section{Concluding Remarks}\label{sec:conc}
We introduced the \rcpc model checking toolkit for verifying and simulating reconfigurable multi-agent system. We formally presented the syntax and semantics of \rcpc language in relation to the \rcp framework~\cite{AbdAlrahmanP21,rcp}, and we used it to model and reason about a nontrivial case study from the realm of reconfigurable and self-organising MAS. Our semantics approach consisted of two types of semantics: structural semantics in terms of automata to recover information about interaction features, and execution semantics based on \rcp. The interaction information recovered in the structural semantics is recorded succinctly in the execution one, and thus permits reasoning about interaction protocols and message exchange. \rcpc is supported with a command line tool, a web editor with syntax highlighting and visualisation.

We integrated \rcpc with \nuxmv to enable \ltl symbolic (bounded) model checking. We showed that this specialised integration provides a powerful tool that permits verifying high-level features such as synchronisations, interaction protocols, joint missions, channel mobility, reconfiguration, self-organisation, etc.

\smallskip

\noindent
{\bf Related works.} We report on closely related model-checking toolkits and their relation to \rcpc.

MCMAS is a successful model checker that is used to reason about multi-agent systems and supports a range of temporal and epistemic logic operators. MCMAS is also supported with ISPL, a high-level input language with semantics based on \emph{Interpreted Systems}~\cite{FHMV95}. The key differences with respect to \rcpc are: (1) MCMAS models are enumerative and are exponentially larger than \rcpc ones; (2) actions in MCMAS are merely synchronisation labels while command labels in \rcpc are predicates with truth values changing dynamically at run-time, introducing opportunistic interaction; (3) lastly and most importantly \rcpc is able to model and reason about dynamic communication structure with message exchange and channel mobility while in MCMAS the structure is fixed.

MTSA toolkit~\cite{DIppolitoFCU08} is used to reason about labelled transition systems (LTS) and their composition according to the simple multiway synchronisation of Hoare's CSP calculus~\cite{Hoare21a}. MTSA uses the \emph{Fluent Linear Temporal logic (FLTL)}~\cite{gianna03} to reason about actions, where a fluent is a predicate indicating the beginning and the end of an action. As the case of MCMAS, the communication structure is fixed and there is no way to reason about reconfiguration or even message exchange.

SPIN~\cite{holzmann1997model} is originally designed to reason about concurrent systems and protocol design. Although SPIN is successful in reasoning about static coordination protocols, it did not expand its coverage to multi-agent system features. Indeed, the kind of protocols that SPIN can be used to reason about are mainly related to static structured systems like hardware and electronic circuits.

Finally, \nuxmv~\cite{CimattiG12} is designed at the semantic level of transition systems. \nuxmv implements a large number of efficient algorithms for verification. This makes \nuxmv an excellent candidate to serve as a backbone for several specialised purpose model-checking tools. For this reason, we integrate \rcpc with \nuxmv.
\smallskip

\noindent
{\bf Future works.}
We plan to integrate \ltal to \rcpc from~\cite{AbdAlrahmanP21}. Indeed,
the authors in~\cite{AbdAlrahmanP21} provide a \pspace algorithm for
\ltal model checking (improved from \expspace in~\cite{rcp}). This way,
we would not only be able to refer to message exchange in logical
formulas, but also to identify the intentions of agents in the
interaction and characterise potential interacting partners.

Moreover, we would like to equip \rcpc with a richer specification language that allows reasoning about the knowledge of agents and the dissemination of knowledge in distributed settings. For this purpose, we will investigate the possible integration of \rcpc with MCMAS~\cite{LQR17} to make use of the specialised symbolic algorithms that are introduced for knowledge reasoning.

%
%
%
\bibliographystyle{splncs04}
\bibliography{biblio}


\end{document}